\documentclass{iaus}
\usepackage{graphicx}

\title[Point X-ray sources in Ellipticals] 
{Point X-ray Sources in Elliptical Galaxies}

\author[Ivanova \& Kalogera]   
{Natalia Ivanova$^1$%
  \and Vicky Kalogera$^1$}

\affiliation{$^1$Physics and Astronomy Department, Northwestern University, 
2145 Sheridan Rd, Evanston IL 60208 \break email: nata@northwestern.edu}

\pubyear{2005}
\volume{230}  
\pagerange{1--4}
\date{19 September, 2005}
\jname{Populations of High Energy Sources in Galaxies}
\editors{Evert J.A. Meurs \& Giuseppina Fabbiano, eds.}
\begin{document}

\maketitle

\begin{abstract}
We analyze the upper-end  X-ray luminosity function
(XLF) observed  in elliptical galaxies for  point sources.
We propose that   the observed XLF is dominated by transient 
BH systems in outburst and  the XLF shape reflects 
the  black hole (BH) mass  spectrum among old X-ray transients. 
The BH mass spectrum -- XLF connection depends on 
a weighting factor that is  related to the transient duty cycle and
depends on the host-galaxy age,  the BH mass and the donor type
(main sequence,  red giant, or white dwarf). 
We argue that the assumption of a  constant duty cycle 
for all  systems leads to results
inconsistent with current observations.  
The type of dominant donors in the upper-end XLF depends 
on what type of magnetic braking operates:
in the case of  ``standard'' magnetic braking, BH X-ray binaries 
with red-giant donors
dominate, and in the case of weaker magnetic braking prescriptions
main sequence donors dominate.  
\end{abstract}
\firstsection 
%
\section{Transient and persistent Black Hole X-Ray Binaries in Ellipticals}
With {\em Chandra} observations, 
elliptical galaxies out to the Virgo cluster have 
now been studied. These observations have revealed a large number  
of  point X-ray sources and early on     
\cite{2000ApJ...544L.101S} identified an
XLF shape that required two power laws with a break at
$\simeq\,3.2\times\,10^{38}$\,ergs\,s$^{-1}$.
Following longer exposures for more
ellipticals  (\cite[Gilfanov 2004, Kim \& Fabbiano 2004]{2004MNRAS.349..146G,
2004ApJ...611..846K}) have confirmed that the combined sample of sources
from all observed galaxies requires two power laws and a break at $5\pm 1.6\times
10^{38}$\ ergs\ s$^{-1}$ and the best-fit slope of the
upper end to be $\alpha_{\rm d}=2.8\pm0.6$.

We analyze the upper-end of XLF assuming:
(i) the XLF above the break at $5\times
10^{38}$ erg s$^{-1}$ is populated by X-ray binaries (XRBs) with black hole 
(BH) accretors (\cite[Sarazin et al. 2000]{2000ApJ...544L.101S}); 
(ii) the vast majority of these BH-XRBs are part
of the galactic-field stellar population in ellipticals; 
(iii) donor masses are lower than $\simeq 1-1.5$\,M$_{\odot}$
(in accordance to the current estimates for the ages of
stellar populations in ellipticals) and
(iv) we adopt  the  current  understanding  
for the  origin  of  transient behavior 
in XRBs  (\cite[King 2005]{BH_book_ch13}). 
We consider typical mass transfer  (MT)  rate  associated  with  each type  of  XRB  donor
($\dot{M}_{i}$) and compare it to  the critical MT rates for transient
behavior  ($\dot{M}_{\rm  crit}$):   if  $\dot{M}_{i}  <  \dot{M}_{\rm
crit}$, the binary  system is  assumed to  be a  transient  X-ray source and
 we assume that during the outburst the X-ray luminosity is equal to the BH accretor Eddington
luminosity.
                                
Mass transfer  in BH  XRBs with low-mass main sequence (MS)  donors is
driven by  angular  momentum losses  due to  magnetic
braking (MB) and gravitational radiation (GR).  
From   detailed  binary
evolutionary  calculations using  the  stellar evolution  and MT  code
(\cite[Ivanova \& Taam 2004]{2004ApJ...601.1058I}), we find
that if MB follows the Skumanich law
(\cite[taken as in Rappaport et al. 1983]{1983ApJ...275..713R}),  
then XRBs with BHs  $\le 10 M_\odot$ are persistent as long
as the  donor masses  are $> 0.3\, M_\odot$. 
The MT rates are about $0.01-0.25$  of the BH's Eddington  rate.  
As a result, the   persistent  X-ray   luminosity   for  
these   systems  is   $\le 10^{38}$\,erg\,s$^{-1}$.  
For  $M_{\rm  BH} > 10$\,$M_\odot$, the outburst luminosity is in excess of
$\simeq  1.5\times10^{39}$\,erg\,s$^{-1}$.  
This  limit
is  comparable to  the highest  luminosity seen  currently in  XLFs of
ellipticals (\cite[Kim \& Fabbiano 2004]{2004ApJ...611..846K}),  
and therefore  these systems
cannot  contribute  significantly  to  the  observed  XLFs. 
Outbursts are possible from  transient BH-MS  with  $M_{\rm BH}  <
10$\,$M_\odot$    and    donors     less    massive    than    $\simeq
0.3$\,M$_\odot$; these  low mass donors are out  of thermal equilibrium. 
In the case of the MB prescription derived for fast rotating systems, which
are BH-MS binaries (\cite[Ivanova \& Taam 2003]{2003ApJ...599..516I}, IT), 
BH-MS systems are transient for all  BHs  masses $M_{\rm  BH}>  3 M_\odot$  
and  for  all low-mass  MS donors. 
If a donor is a low-mass subgiant or red giant (RG),  
it has been shown (\cite[King 2005]{BH_book_ch13})  
that  such  XRBs are transient, regardless of the BH mass. 
The evolution of mass-transferring BH systems with a white dwarf (WD) very weakly depends on the
BH mass. A typical life-time of such a system in the persistent stage is
$\sim 10^7$ years and only during  $10^6$ years a BH-WD will have MT rates 
in excess of the Eddington limit.
We  conclude  that  all BH  binaries  that contribute  to  the  current
upper-end   XLFs  of   ellipticals  are expected  to   be  transient
sources.
\section{Weighting Factor for the BH Mass Spectrum  and the Duty Cycle}
The  transient BH XRBs (those in outburst)  contributing to the upper-end XLF 
is  a sub-set of the true  population of BH
XRBs  in ellipticals  determined by  the duty  cycle of  BH transient
binaries.  For the  general case  of a  transient duty  cycle  that is
dependent on the BH accretor mass, the
differential  XLF   $n(L)_{\rm  obs}$  and  the   underlying BH  mass
distribution in XRBs $n(m)_{\rm BH}$ are connected by:
 \begin{equation}
n(L_{X})_{\rm obs} = n(m_{\rm BH}) \times W(m_{\rm BH}), 
 \end{equation}
 where $W(m_{\rm BH})$ is a weighting factor related to the dependence 
of the transient duty cycle on $m_{\rm BH}$. 
The observed  slope of the differential upper-end  XLF is $\alpha_{\rm
d}=2.8\pm0.6$:  $n(L_{X})_{\rm  obs}\propto\,L_{X}^{-\alpha_{\rm  d}}$.
   Assuming  that   $n(m_{\rm
BH})\propto\,m_{\rm  BH}^{-\beta}$  and $W(m_{\rm  BH})\propto\,m_{\rm
BH}^{-\gamma}$,  the  slope  characterizing  the  underlying  BH  mass
distribution in XRBs is 
$ \beta = \alpha_{\rm d} - \gamma $.

At present there are no strong constraints on the duty
cycles either from observations or from theoretical considerations. 
Among known Galactic X-ray transients, typical duty cycles of a few\,
\% is favored for hydrogen donors (Tanaka \& Shibazaki 1996) and
there are no data on duty cycles for transients  with a WD companion. 
For   the   standard   assumption   of  a   constant   duty   cycle,
$\beta=\alpha_{\rm  d}= 2.8\pm0.6$. 
However, according to  binary  population synthesis  models for  the
Milky Way published  so far, the number of  formed BH-WD LMXBs exceeds
the   number  of   BH-MS   LMXBs  by  a   factor  of   100
(\cite[Hurley et al. 2002]{2002MNRAS.329..897H}). If the duty cycle were
similar for both types of systems, 
we would observe a few hundreds BH-WD binaries; 
however none such binaries have been identified.
We therefore investigate how plausible duty cycle
assumptions affect the upper-end XLF shape,
considering a variable (dependent on MT rates) duty cycle
equal to 
$\eta=0.1\times {\dot M_{\rm  d}}/{\dot M_{\rm crit}}$.
This  specific choice of the  dependence is  not solidly
motivated,  however,  it implies a  correlation of the duty  cycle with
how  strong a  transient the  system is:  the further  away  from the
critical MT rate, the smaller the  duty cycle.

\begin{figure}
\begin{center}
{\includegraphics[height=1.5in,width=2.63in]{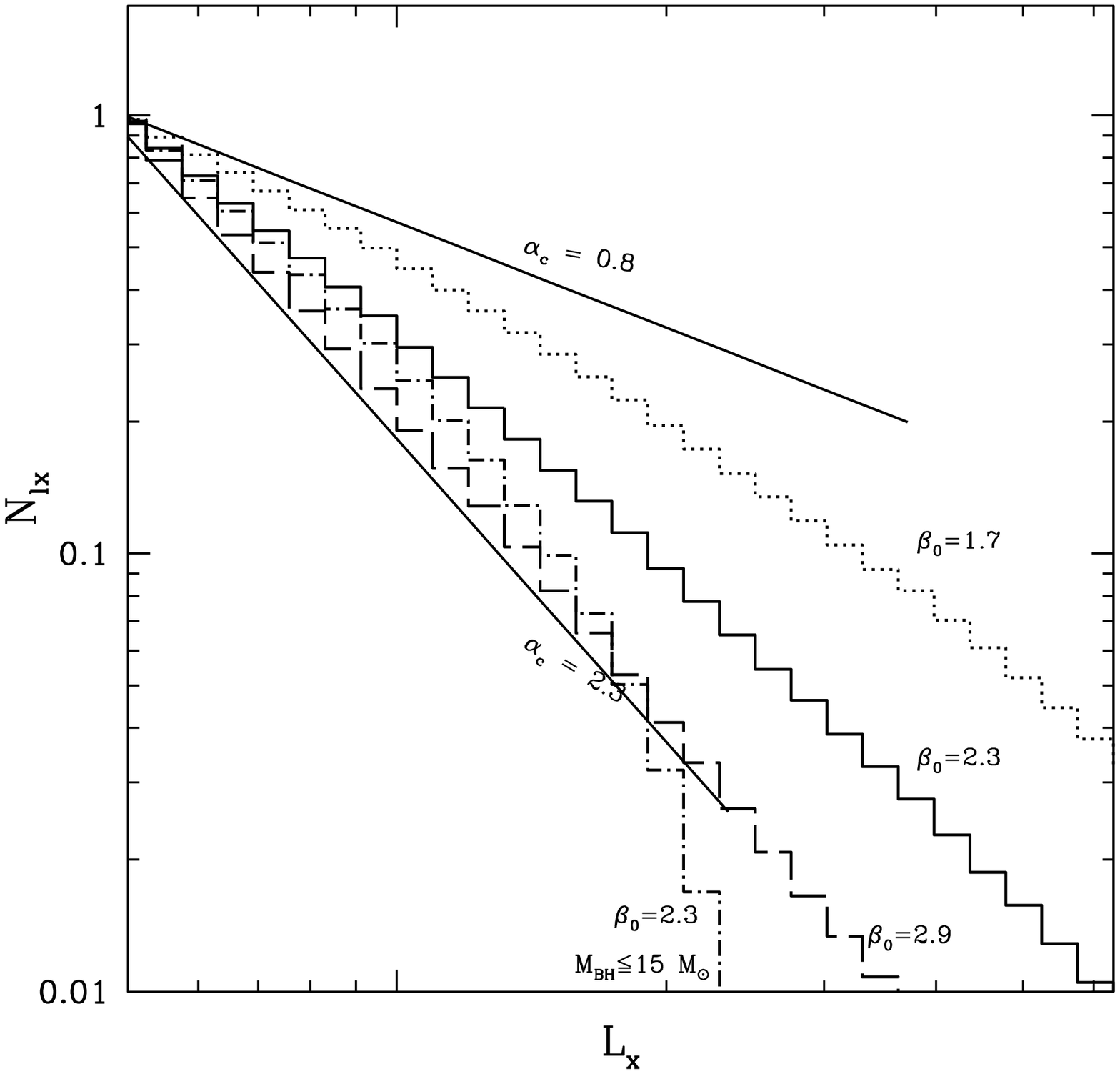}
\includegraphics[height=1.5in,width=2.63in]{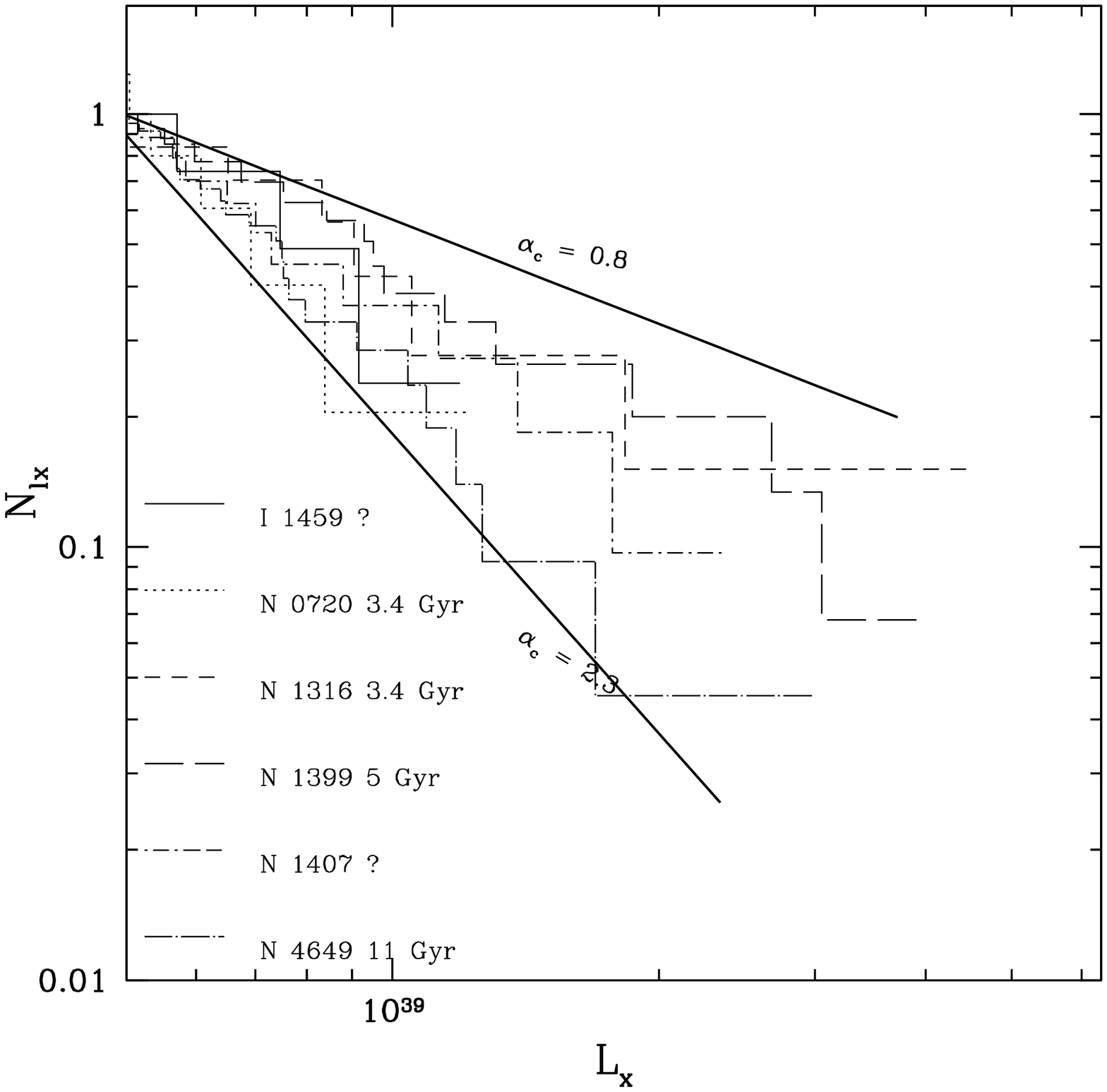}}
\end{center}
\caption{ The left panel shows  model XLFs for BH-MS binaries.  Lines
shows results  for initial $\beta_0=1.7,2.3,2.9$  at an age of  10 Gyr
(dotted, solid  and dashed lines).  The dash-dotted line
is for a  $15 \,M_{\odot}$ BH mass cut-off  with $\beta_0=2.3$.  Thick
solid   lines   corresponds  to   slopes   $\alpha_{\rm  c}=0.8$   and
$\alpha_{\rm c}=2.3$.   The right panel  shows XLFs in  observed early
type galaxies (data are from \cite[Kim \& Fabbiano 2004]{2004ApJ...611..846K},
the ages  of the ellipticals  are from
\cite[Ryden et al. 2001]{2001MNRAS.326.1141R} and \cite[Temi et al. 2005]{2005ApJ...622..235T}).  }
\label{xlf_art}
\end{figure}

The expression for $\eta$ in the case of WD and RG donors can be found analytically.
For RG we obtain (using MT rates for RGs from 
\cite[Webbink et al. 1983]{1983ApJ...270..678W})
${\dot m_{\rm RG}}/{\dot m_{\rm crit}} \simeq 0.2 a^{-0.7} m_{\rm d}^{1.5}$.
Therefore,  for   RG  donors,   $\gamma=0$.   Instead   there  is
a significant dependence  on the RG  donor mass.  Since  in ellipticals
the typical mass  for RG donors is  about the same as the  mass of the
turn-off of  MS stars,  the duty  cycle for RG  donors depends  on the
turn-off mass, and hence on the age T [Gyr] of the elliptical.
Assuming a flat in the logarithm distribution  of orbital  separations before  MT starts, we find:
 \begin{equation}
W(T) \simeq 0.03  \times T^{-0.5}\, . 
\label{w_for_rg}
 \end{equation}

In order to find  the  probability  that  a  BH-WD  
contributes to  the upper-end XLF at  an elliptical age, we assume
that  (i) all  accreting  BH-WD  systems were  formed  within a  short
interval of elliptical ages 
several  Gyrs ago; and (ii) the binary is
a transient at time $T$.  We further 
adopt  a constant  BH-WD formation  rate during this interval.  We obtain:
 \begin{eqnarray}
W(T;m_{\rm BH}) & = & 1.6 \times 10^{-4} m_{\rm BH}^{-0.5} t_1^{-7/4}  \frac{1 - (t_2/t_1)^{-3/4}} { 1 - (t_2/t_1)}
\label{w_for_bhwd}
 \end{eqnarray}
Here $t_1$ and $t_2$ are the time in Gyrs from the current moment to the
start and the end of BH-WD binaries formation.
Therefore for WD donors $\gamma=0.5$ and $W$  depends on both the BH mass and the galaxy age.

As  discussed previously, for  IT MB,  a BH-MS  system  is transient
throughout the  MT phase.  It also can be seen that the MT dependent duty cycle
is about few \%, consistent with the observations.
Prolonged  mass accretion onto the BHs  affects their
mass spectrum.  Since this effect cannot be  included analytically, we
have examined it quantitatively  using simple Monte Carlo simulations.
We set up the simulations assuming  a flat BH-MS birth (MT onset) rate
and a flat mass distribution for  donors at the onset of the MT phase,
without any restrictions on the mass ratio $q_{\rm BH}$. For the MT  evolution we took
into  account both IT  MB and  GR, and  if
the MT  timescale is longer than  the thermal timescale  of the donor,
the donor radius evolution is simply proportional to the mass lost due
to MT.  On the  other hand, if  the MT  timescale is shorter  than the
donor's  thermal   timescale,  the  donor   is  out  of   the  thermal
equilibrium. In this case we  modify the evolution of the donor radius
using a prescription that is  in acceptable agreement with our results
from detailed  MT calculations with  the stellar evolution:  $\delta r
\sim \delta m \sqrt{\dot m_{\rm TH} / \dot m}$, where $\dot m_{\rm TH}
= m_{\rm d}/ t_{\rm TH}$ is  the MT rate driven on the donor's thermal
timescale $t_{\rm TH}$. 

Based on the results of our  Monte Carlo simulations we find that: (i)
due to accretion the BH  mass spectrum slope increases by about $0.2$,
i.e., $\beta  = \beta_{\rm  0}+0.2$, where $\beta_{\rm  0}$ is  the BH
mass slope at MT onset; (ii) the  slope of the BH mass spectrum at the
beginning  of  mass transfer  best  reproduces  the observations  with
$\beta_{0}=2.3\pm0.6$ (see Fig.~1).  We also find that the
relation between $\beta$  and $\beta_{0}$ is not sensitive  to the age
of the elliptical.
\section{Conclusions}
We conclude that all BH binaries contributing to the upper-end XLF of ellipticals are transient.
A constant transient duty cycle independent of the donor type can be excluded
unless BH-WD transients have very weak outbursts and can not be detected.
The upper-end XLF is formed by an 
underlying mass spectrum of accreting BHs.
The BH X-ray transients have a dominant 
donor type and an accreting BH mass spectrum slope $\beta$
that depends on the strength of MB angular momentum loss. 
In particular,
in the case of Skumanich-type  MB, 
the XLF is dominated by BH-RG binaries and $\beta=2.8\pm0.6$.
In the case of MB for fast-rotators, 
only BH-MS transients significantly contribute to 
the upper-end XLF and $\beta=2.5\pm0.6$ \ ($\beta_0=2.3\pm0.6$).
If the relative fraction of BH-RG transients in ellipticals is larger 
than the observed relative fraction in our Galaxy, we expect that 
the BH-RG binaries contribution will lead to a time-dependence of 
XLF slopes, where younger ellipticals will have a  slope 
predicted for BH-RG binaries, and older ellipticals 
a steeper slope predicted for BH-MS binaries.
\begin{acknowledgments}
This work is partially supported by a {\em Chandra}  Theory Award  
to N.\ Ivanova  and a  Packard Foundation
Fellowship  in  Science  and  Engineering  to V.\  Kalogera.
\end{acknowledgments}

\begin{discussion}
\discuss{Lipunov}{The initial mass function can 
not be described as simple power law.}
\discuss{Ivanova}{It is not assumed or insisted.
This is what the XLFs tell us about how the BHs mass spectrum 
looks like, for a specific type of donors.
It does not represent the whole spectrum of all 
ever formed BHs, only for selected type of donors.
We hope that this result will
provide additional constrains for these 
scientific teams who work on the connection between
pre-collapse stars and the resulting masses of formed compact objects.}
\discuss{Kim}{Can you constrain the minimum mass of BH?}
\discuss{Ivanova}{No, we can not make it directly. 
We can only give constrains on the BH mass spectrum 
given by our understanding of their donors is correct.
E.g., we can not rule out BH less massive than $3 M_\odot$ as 
the possible lack of BH-MS binaries with BHs of such masses in XLFs 
can be the result of different selection effects originating in
the  binaries evolution, but not in BHs formation.
}
\discuss{Sivakoff}{You predict that all XRB in the high end of the XLF are
transient. Can you comment on the time-scales expected for this transience?
In particularly, are the time-scales observable on our life-times?}
\discuss{Ivanova}{We can only give estimates for the duty cycles, but not for
the outbursts durations. The latter will have to come from observations.}
\end{discussion}

\end{document}